\documentclass[5pt]{article}

\usepackage[letterpaper]{geometry}
\usepackage{spconf,amsmath,epsfig,float, url, bm}
\usepackage{enumitem}

\usepackage{fancyhdr}
\newcommand{\norm}[1]{\left\lVert#1\right\rVert}

\begin{document}\sloppy

\def\x{{\mathbf x}}
\def\L{{\cal L}}

\title{Fast block structure determination in AV1-based multiple resolutions video encoding}
%
\name{Bichuan Guo$^{\ast}$, Yuxing Han$^{\dag}$, Jiangtao Wen$^{\ast}$\thanks{This work was supported by the Natural Science Foundation of China (Project Number 61521002).}}
\address{$^{\ast}$Tsinghua University, $^{\dag}$South China Agricultural University  \\
jtwen@tsinghua.edu.cn}
%
%
%

\maketitle

\begin{abstract}
The widely used adaptive HTTP streaming requires an efficient algorithm to encode the same video to different resolutions.
In this paper, we propose a fast block structure determination algorithm based on the AV1 codec that accelerates high resolution encoding, 
which is the bottle-neck of multiple resolutions encoding.
The block structure similarity across resolutions is modeled by the fineness of  frame detail and scale of object motions,
this enables us to accelerate high resolution encoding based on low resolution encoding results.
The average depth of a block's co-located neighborhood is used to decide early termination in the RDO process.
Encoding results show that our proposed algorithm reduces encoding time by 30.1\%-36.8\%, while keeping BD-rate low at 0.71\%-1.04\%. Comparing to the state-of-the-art, our method halves performance loss without sacrificing time savings.
\end{abstract}

\section{Introduction}
\label{sec:intro}

Adaptive HTTP streaming has now been widely used by most video content providers
to improve the quality of experience (QoE) of their users \cite{Oyman2012QoE}.
Videos are stored as multiple representations with varying sizes and qualities,
and the client-side player requests a suitable representation
according to the network condition \cite{Li2003}.
The video spatial resolution, being one of the greatest factors affecting the video bit-rate,
is used by multiple popular video-sharing websites (e.g. Youtube, Twitch) as the primary option to control the video quality.
Therefore, a fast algorithm that encodes the same video to different resolutions is certainly of great interest.

AOMedia Video 1 (AV1) \cite{av1} is an emerging video codec developed by the Alliance for Open Media,
which is open-source and royalty-free.
Comparing to its predecessor VP9, 
it offers numerous new coding tools to achieve a cutting edge coding efficiency,
at the expense of complexity.
As a result, fast encoding is very challenging especially regarding to adaptive HTTP streaming, 
where the same video needs to be encoded for multiple times.

To speed up the encoding, 
one needs to exploit the correlation between multiple rate distortion optimizations (RDO).
The block structure is mainly determined by the fineness of frame detail,
and the amount of inter-frame motion. 
Therefore, the block structure preserves to a large extent during target resolution rescaling \cite{Schroeder2015}.
Motion vectors can also be reused, as they represent the motion of objects and 
therefore shall be consistent across different resolutions up to a rescale factor.
However, an early termination algorithm that prevents unnecessary block structure search would also
terminate all subsequent motion estimations, 
hence it is not surprising that the encoding complexity can be greatly reduced by solely considering block structures, as proven by several papers \cite{Schroeder2017}\cite{Praeter2015}.

In this paper, we propose a fast block structure determination algorithm for AV1-based multiple resolutions encoding.
Encoding processes with high target resolutions are accelerated by referring to a low target resolution encoding process to infer block structures and execute RDO early termination accordingly.
Based on our statistical analysis, 
the average block depth of the co-located neighborhood, 
coupled with adaptive search range and threshold value,
yield early termination decisions according to two distinct strategies designed for different block sizes.
 
The rest of the paper is organized as follows. 
Related work is presented in Section 2. 
The block structure of AV1 and the block structure similarity across resolutions are studied in Section 3.
Section 4 describes how to exploit this cross-resolution similarity to gain information about block structures based on low resolution RDO results,
and in turn accelerates high resolution encoding.
Experimental results are given in Section 5, 
and Section 6 concludes the paper.

\section{Related Work}

There are many frameworks dedicated to reduce the overall complexity of multiple representations video encoding.
Transcoding methods \cite{Ahmad05}\cite{Chen15} reduce the complexity by reusing RDO mode decisions and motion estimation results,
obtaining a new representation through residual re-quantization from an existing bit-stream. 
However this introduces significant quality losses due to re-quantization,
and it is not suitable for multiple resolutions encoding.
Scalable video coding \cite{Schwarz07} implements multiple layers in a single bitstream corresponding to different qualities,
so that users can adaptively request the suitable sub-stream. 
While its performance is better than transcoding, 
it is still not desirable comparing to single layer coding,
especially when the number of layers is high,
and where bandwidth resource is limited.

The idea of fast multiple representations encoding was first introduced in \cite{Zaccarin2002}.
The RDO redundancy among different encoding processes is examined in order to gain speedup without introducing substantial rate-distortion (RD) loss.
\cite{Finstad11} proposed a preliminary framework for same resolution, multiple target bit-rates encoding, 
where RDO decisions from the highest bit-rate encoding process were copied to other processes.
This method resulted in considerable RD loss and was later improved by an RDO tree pruning algorithm \cite{Schroeder2015}\cite{Schroeder2017}.
\cite{Schroeder20152} proposed a heuristic multiple resolutions encoding algorithm for HEVC,
where low resolution encoding is accelerated by a high resolution reference encoding.
However, multiple resolutions encoding are often done in parallel,
and the highest resolution takes the longest time.
Therefore, the aforementioned approach does not offer much benefit in practical situations.

As its main contribution, this paper proposes a fast block structure determination algorithm for AV1 that 
accelerates high resolution encoding based on low resolution encoding results.
It has perfect parallel compatibility, 
in a sense that when all representations are encoded simultaneously,
the time reduction on high resolution encoding fully translate into overall time reduction.
Unlike most empirical algorithms listed above,
the block structure similarity across resolutions is modeled by the fineness of frame detail and scale of object motions, 
the determination algorithm is then derived from statistical hypothesis testing.

\section{Block structure similarity}

\subsection{Block structures in AV1}
\begin{figure}[t]
\begin{center}
\includegraphics[width=0.3\textwidth]{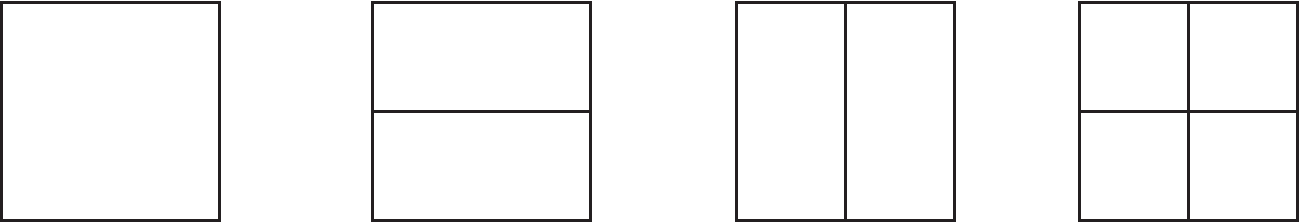}
\end{center}
\caption{\label{fig:block-partition}%
Four ways to partition a square block in AV1.}
\end{figure}

Similar to HEVC, AV1 also uses a quadtree-based block structure.
Each frame is partitioned into 64$\times$64 blocks, 
and each square block can be further partitioned in a recursive manner.
To be specific, a $2N$$\times$$2N$ block can be partitioned into four $N$$\times $$N$ blocks (4-way split), 
two $N$$\times$$2N$ blocks (2-way horizontal split), two $2N$$\times$$N$ blocks (2-way vertical split), or remain unpartitioned
(see Fig.~\ref{fig:block-partition}).
The smallest block size is 4$\times$4, and non-square blocks cannot be further partitioned.
There are experimental tools in AV1 that support 2$\times$2 blocks and the so-called ``T-split'', however they will not be discussed in this paper as they are not included in the standard settings. 

As there are square and non-square blocks, 
our following discussion will categorize 2-way splits as non-split, 
since they both share the property that cannot be further partitioned. 
This simplification allows us to only talk about non-split and 4-way split as traditional block-based coding formats.
To avoid ambiguity, the depth of a block will refer to its longer edge, 
starting from zero, i.e. the depth of a $w\times h$ block is
\begin{align*}
	d = \min(\log_2(64/w), \log_2(64/h)),
\end{align*}
therefore a 64$\times$64 block's depth is 0, a 32$\times$16 block's depth is 1, and so on. The maximum depth is 4.

\subsection{Cross-resolution similarity}

\begin{figure}[t]
\begin{center}
\includegraphics[width=0.4\textwidth]{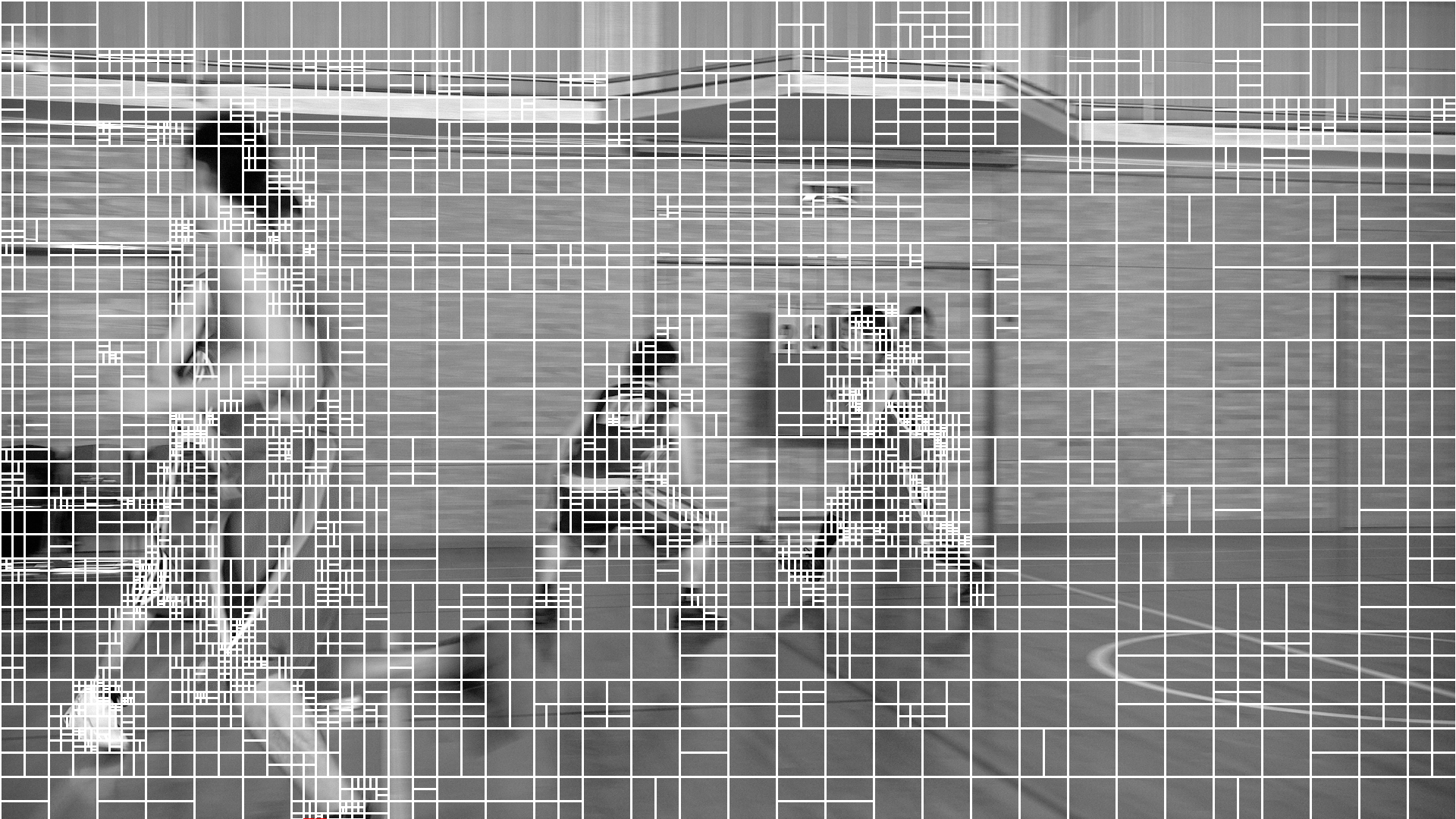} \\
(a) 1920$\times$1080\\
\includegraphics[width=0.4\textwidth]{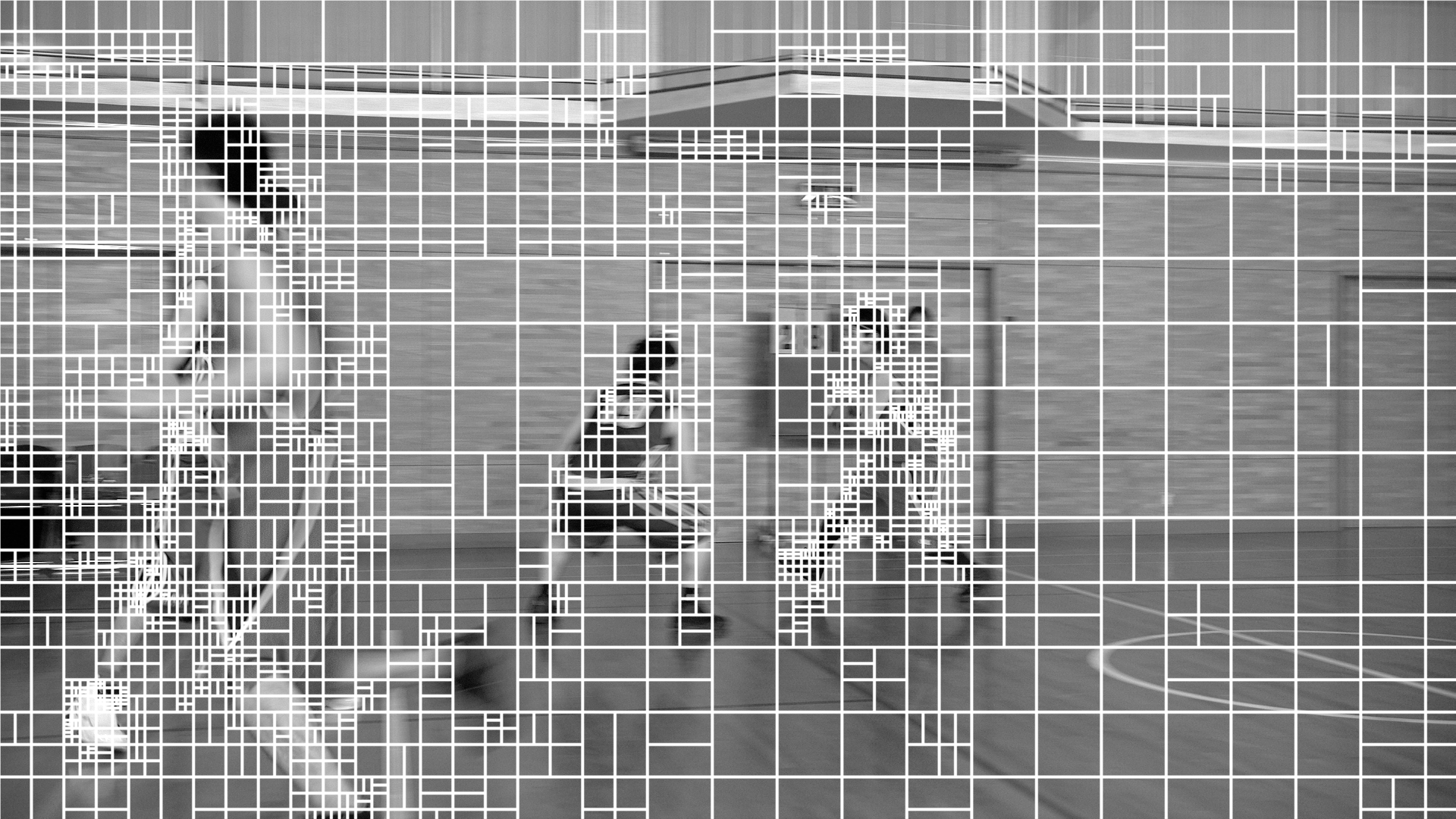} \\
(b) 1440$\times$810
\end{center}
\caption{\label{fig:block-similarity}%
Block structure of the 1st frame of the \textit{BasketballDrive} sequence, QP 27}
\end{figure}

To observe the block structure similarity among different resolutions,
the \textit{BasketballDrive} test sequence from CTC \cite{CTC} is encoded using the AV1 codec to target resolutions 1920$\times$1080 and 1440$\times$810, both using a constant QP of 27.
Fig.~\ref{fig:block-similarity} shows the block structure of the first frame. 
It is clear that finely partitioned areas are in common:
the athletes and the ceiling, i.e. areas with high level detail or large motion.
This inspires us to use the fineness of frame detail and the scale of object motions to model the block structure similarity across resolutions.

Therefore, an assumption is made that there exists a continuous function $f$ defined on the frame plane which
represents the fineness of frame detail and scale of object motions at each point.
The value of this function is inherent to the scene, and does not depend on the target resolution of the encoder.
The closed form of $f$ will most likely depend on the image gradient and the optical flow,
however, our following discussion only relies on the existence of such a resolution-invariant continuous function.

Denote the high (low) resolution encoding process by $P_1$ ($P_2$). Suppose a block $B_1$ with depth $d$ in $P_1$ has a neighbor block $B_2$ with depth $d$ in $P_2$ (that is, they are close to each other).
Let $X_i$ denote the partition choice of $B_i$, such that $X_i=1$ when $B_i$ is 4-way split, otherwise (non-split or 2-way splits) $X_i=0$.
Our previous observation states that $X_i$ tends to be 1 where $f$ takes large values, thus an assumption is made that the partition choice of a block is decided by its size and the value of $f$ in its area.
Formally, denote the average value of $f$ in $B_i$ as $\mu  B_i$, $X_i$ can be modeled as a Bernoulli random variable with
$P(X_i=1) = g_i(\mu B_i)$,
where $g_i$ is a monotonically increasing function,
representing the positive correlation between the tendency of any block in $P_i$ with depth $d$ being partitioned,
and the value of $f$ in the block's area.
The assumption of $f$ being resolution-invariant and continuous gives the following relation:
\begin{align}
	EX_1 = g_1(\mu B_1) &\approx g_1(\mu B_2) \nonumber \\
	& =g_1\circ g_2^{-1}(EX_2). \label{eqn:ex1-ex2}
\end{align}
The approximation is due to $B_1$ being close to $B_2$, therefore the average values of $f$ are also close.
It then follows that $g_1\circ g_2^{-1}$, the composition of $g_1$ and the inverse of $g_2$, is also increasing.
This explains our intuition that wherever blocks in $P_2$ are finely partitioned, so are those in $P_1$.

However, there are many cases where blocks in Fig.~\ref{fig:block-similarity}(b) are not partitioned, 
but their neighbor blocks in Fig.~\ref{fig:block-similarity}(a) are.
This inconsistency results from the variance of $X_i$.
In fact, the  block structure similarity is more consistent in the average sense.
Suppose $B_i$ has a neighbor of blocks $B_{ij}$ $(1\le j\le n)$ in $P_i$ with the same size as $B_i$,
relate $X_{ij}$ to $B_{ij}$ as $X_i$ to $B_i$, define 
\begin{align}
	X_i' = \frac{1}{n}\sum_j X_{ij}.
\end{align}
Assuming independence between $X_{ij}$, we have
\begin{align}
	EX_i' &= \frac{1}{n}\sum_{j=1}^n g_i(\mu B_{ij}), \label{eqn:exi-prime}\\
	\sigma X_i' &= \frac{1}{n}\sqrt{\sum_{j=1}^n g_i(\mu B_{ij})(1-g_i(\mu B_{ij}))}. \label{eqn:sigmaxi-prime}
\end{align}

If the neighborhood area is sufficiently small, 
$\mu B_{ij}$ is close to $\mu B_{i}$. 
Then $EX_i'\approx EX_i$, 
$\sigma X_i'\approx \frac{1}{\sqrt{n}}\sigma X_i$.
Therefore the relation in (\ref{eqn:ex1-ex2}) also holds for the neighborhood of $X_i$ in place of $X_i$ itself, 
with smaller variance.
If all partitions beyond depth $d+1$ is ignored, 
$X_i'+d$ is essentially the average block depth among $B_{ij}$.
This suggests a more consistent similarity
between the average block depth of co-located neighborhoods.

\section{Block structure inference model}

\subsection{Fast block structure determination}
Multiple resolutions encoding are often done in parallel,
where the same video is encoded to different resolutions on a multi-core server,
and the overall time cost depends on the most time-consuming encoding process, 
i.e. the one with the highest target resolution.
As a result, a fast block structure determination algorithm that
accelerates high resolution encoding also reduces overall time cost by the same amount.

The block structure similarity among co-located neighborhoods,
as proven by (\ref{eqn:ex1-ex2}),
can be exploited to
provide useful information about block structures for high resolution encoding.
Using the same terminology,
the objective can be stated as follows.
It is to be decided whether $B_1$ should be partitioned,
which relies on the distribution of $X_1$, a function of $EX_2$.
Since $P_2$ takes shorter time than $P_1$,
the partition results near $B_2$ should be available,
i.e. for each $B_{2j}$ the observed partition result is $\tilde{X}_{2j}$,
hence the observed value of $X_2'$, which is the sample mean $\tilde{X}_2'$, is always available.

Now, (\ref{eqn:exi-prime})(\ref{eqn:sigmaxi-prime}) show that  $EX_2 $ can be approximated with the sample mean $\tilde{X}_2'$ while keeping variance small.
By (\ref{eqn:ex1-ex2}),  $g_1\circ g_2^{-1}$ also needs to be determined,
however this can only be done with history statistics,
in other words, the value of  $X_1$ is needed.
In a fast determination algorithm, instead of running full RDO in $P_1$,
the encoding results from $P_2$ is used to make inference,
which may result in suboptimal decisions and makes obtaining $X_1$ impossible.

\begin{figure}[t]
\begin{center}
\includegraphics[width=0.45\textwidth]{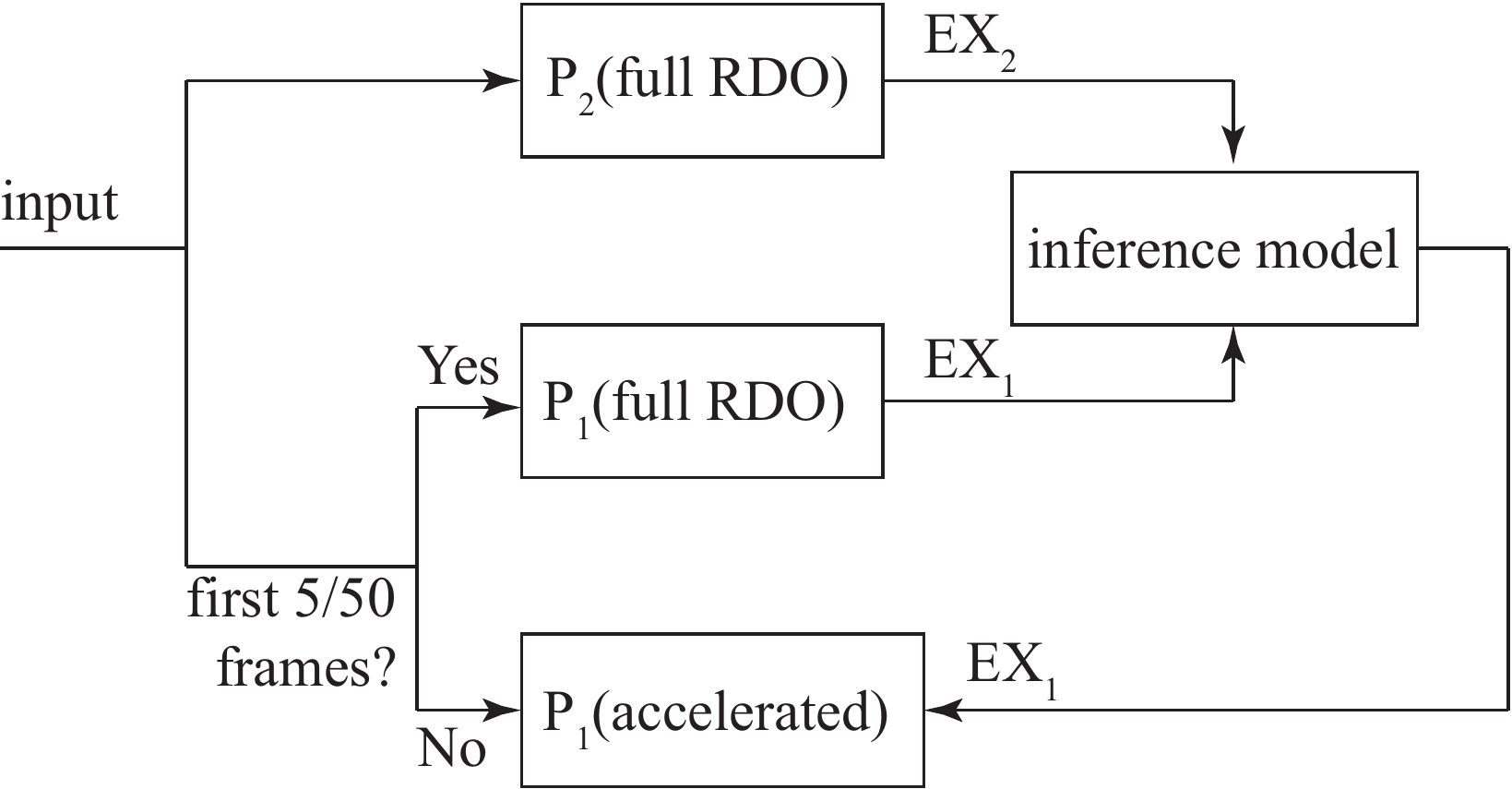}
\end{center}
\caption{\label{fig:procedure}%
The procedure of accelerated encoding.}
\end{figure}

Therefore,  the fast determination algorithm is occasionally disabled
to evaluate $X_1$. Coupled with $\tilde{X}_2'$ which is always available,
the information about $g_1\circ g_2^{-1}$ can be obtained by (\ref{eqn:ex1-ex2}).
Once $g_1\circ g_2^{-1}$ is determined, the fast determination algorithm is relatively simple:
compute $EX_1$ with (\ref{eqn:ex1-ex2}), if it is sufficiently small,
i.e. $P(X_1 = 0)$ is sufficiently large,  the RDO is terminated
so that $B_1$ remains non-split without traversing all its partition possibilities.
In this way, the RDO is shortened and time saving is achieved.
In our implementation, for every 50 frames, the first 5 frames are encoded without the fast determination algorithm.
This ensures that the majority of frames (90\%) are encoded with acceleration, 
and the statistical model is always up to date. 
Fig.~\ref{fig:procedure} shows the procedure of our accelerated encoding system.
For every 50 frames, the first 5 frames are fully encoded both in $P_1$ and $P_2$, 
which give observed value of $X_1$ and $\tilde{X}_2'$, respectively. 
They are used to update the inference model described in the next subsection.
For the rest of the frames, the updated inference model, together with estimates of $EX_2$ from $P_2$, 
yield estimates of $EX_1$ for fast block structure determination.

\subsection{The inference model}

It remains to be shown how to choose a neighborhood with suitable size to compute $\tilde{X}_2'$, 
how to deal with $g_1\circ g_2^{-1}$, and how the algorithm affects performance.

The algorithm may fail in two scenarios:
(i) $B_1$ should not be partitioned but was given a high $EX_1$ estimate,
unnecessary RDO quadtree search is conducted which increases time cost;
(ii) $B_1$ should be partitioned but is given a low $EX_1$ estimate, 
its subsequent RDO is terminated and results in RD loss.
If ``$B_1$ should not be partitioned'' is our null hypothesis,
(i) is the type I error, and (ii) is the type II error.
In practical situations, the resulting RD loss needs to be limited,
while reducing the time cost as much as possible.
Therefore, a type II error rate threshold $\epsilon$ is set, 
an algorithm that has a type II error rate smaller than $\epsilon$,
and minimizes the type I error rate is ideal for our purpose.
A smaller $\epsilon$ reduces the type II error, improving the RD performance,
but in turn increases the type I error and reduces time savings, and vice versa.
Therefore, $\epsilon$ can be used to control the trade-off between acceleration and RD performance.

According to the previous section, our criteria for terminating $B_1$'s RDO is whether the estimated $EX_1$ is small.
Therefore the inference model also includes a threshold $\tau$ that decides if $EX_1$ is sufficiently small.
Since $EX_2$ is used to compute $EX_1$ by (\ref{eqn:ex1-ex2}),
and $g_1\circ g_2^{-1}$ is monotonically increasing,
the criteria can instead be defined based on $EX_2$, i.e. if $EX_2 < \tau$,
the predicted value of $X_1$ (denote by $\hat{X}_1$) is 0, and vice versa.
The type I error occurs when $\hat{X}_1 = 1$ and the observed value $\tilde{X}_1 = 0$, and the type II error occurs when $\hat{X}_1 = 0$ and $\tilde{X}_1 = 1$.

The type I/II errors come from two factors, the randomness of $X_1$, and the inaccuracy in estimating $EX_2$.
It is thus important to choose a suitable sized neighborhood to provide accurate and consistent estimation of $EX_2$.
To this end, a few more assumptions are needed to analyze (\ref{eqn:exi-prime}).
Fix $B_2$, denote the vector from $B_2$ to $B_{2j}$ by $\bm \eta_j$,
assume $\mu B_{2j}$ to be continuous as $B_{2j}$ moves on the frame plane, it is natural to model it with a generalized two dimensional Wiener process:
\begin{align} \label{eqn:wiener}
	\mu B_{2j} \sim \mathcal{N}\bigg(\mu B_2 + \bm \beta \cdot \bm \eta_j, \norm{\bm \eta_j}\sigma^2\bigg),
\end{align}
where vector $\bm \beta$ represents the deterministic drift, and $\sigma^2$ represents the uncertainty. 
Furthermore assume that $g_2$ is differentiable and has a first order Taylor expansion
at $\mu B_2$:
\begin{align} \label{eqn:taylor}
	g_2(\mu B_{2j}) = g_2(\mu B_2) + g_2'(\mu B_2)(\mu B_{2j} - \mu B_2),
\end{align}
by (\ref{eqn:exi-prime})(\ref{eqn:taylor}), for the sample mean $\tilde{X}_2'$, which is our estimator of $EX_2$, its bias satisfies
\begin{align}
	|E\tilde{X}_2' - EX_2| &= \bigg|\frac{1}{n}g_2'(\mu B_2) \bm \beta \cdot \sum_j \bm \eta_j\bigg| \nonumber \\
	&\le \frac{1}{n}|g_2'(\mu B_2)| \norm{\bm \beta} \sum_j \norm{\bm \eta_j}. \label{eqn:exi'}
\end{align}
From (\ref{eqn:sigmaxi-prime})(\ref{eqn:exi'}), there is a bias-variance trade-off in choosing the size of the neighborhood.
For if a large neighborhood of $B_2$ is chosen, 
the number of neighbor blocks $n$ that
have the same size as $B_2$ increases, which gives a smaller variance due to (\ref{eqn:sigmaxi-prime}).
However, a large neighborhood also cause $\norm{\bm \eta_j}$ to be large, 
resulting in a large bias upper bound by (\ref{eqn:exi'}).
It is also noteworthy that by (\ref{eqn:wiener}), a large $\norm{\bm \eta_j}$ causes the variance of $\tilde{X}_{2j}$ to increase, 
offsetting (\ref{eqn:sigmaxi-prime}). 
To sum up, there is a best neighborhood size that balances the bias and variance of the estimator $\tilde{X}_2'$, neither too large nor too small.

The discussion above assumed that a large neighborhood always leads to a large $n$ in (\ref{eqn:sigmaxi-prime}).
This is not always the case, especially when the depth of $B_2$ is large,
as most of its neighbor blocks remain non-split at lower depths.
Also, when $n$ is small, the spatial distribution of $B_{2j}$ is less likely to be balanced around $B_2$,
in the sense that the vector sum $\sum \bm \eta_j$ in (\ref{eqn:exi'}) is less likely to cancel out.
In this case, a large neighborhood significantly increases the bias, and fails to reduce the variance.
As a result, for small sized blocks, $\tilde{X}_2$ itself rather than $\tilde{X}_2'$ is used to estimate $EX_2$ (equivalently, the neighborhood size is set to zero).
In fact, small blocks are so unlikely to be partitioned,
that the number of type I errors often vastly exceeds the number of type II errors.
In this case, the neighborhood size is chosen to minimize the total number of type I and II errors, without restricting the type II error rate.

\subsection{Implementation}
\begin{figure}[t]
\begin{center}
\includegraphics[width=0.18\textwidth]{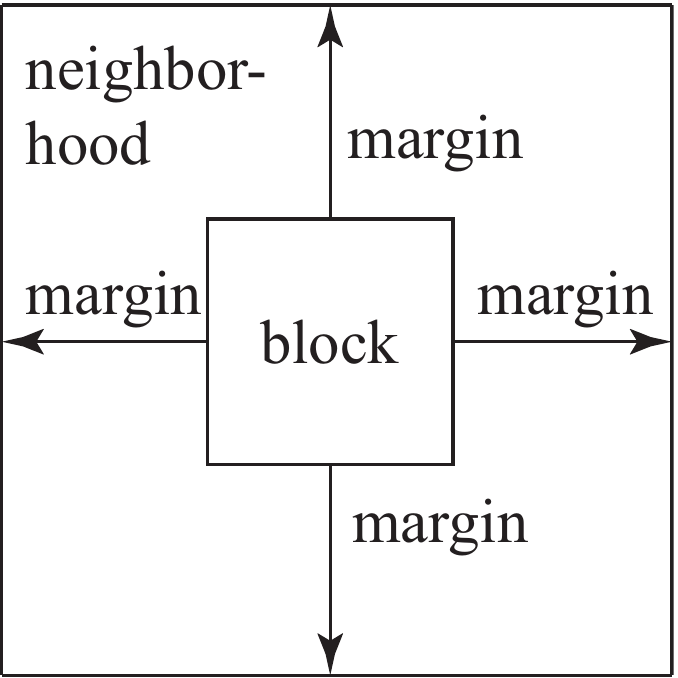}
\end{center}
\caption{\label{fig:neighbor}%
The neighborhood of a block.}
\end{figure}

In our implementation, the neighborhood of a square block is a square formed by
expanding the block's edge in each direction by a specified margin, see Fig.~\ref{fig:neighbor}.
Based on our previous analysis, the implementation of our fast block structure determination algorithm is described as follows.

For every 50 input frames, the first 5 frames are both fully encoded in $P_1$ and $P_2$.
\begin{itemize}
\item For each $d\in \{0,1,2\}$, all square blocks with depth $d$ (regardless of further partition) is found in $P_1$($P_2$), denote them by $\Omega_1$($\Omega_2$). For any block $B_1$ in $P_1$, $\tilde{X}_1$ is the partition choice of $B_1$, and $\tilde{X}_2'$ is the average block depth of $\Omega_2$ within the co-located neighborhood, ignoring further partitions beyond depth $d+1$. The best margin and the corresponding $\tau$ are searched from a discrete set of values. The threshold $\tau$ is chosen as the largest value as long as the type II error ($\tilde{X}_2'<\tau,\tilde{X}_1=1$) rate does not exceed the error rate threshold $\epsilon$, in this way the type I error ($\tilde{X}_2'\ge\tau,\tilde{X}_1=0$) rate is automatically minimized. The best margin is then the one that gives the smallest type I error rate.
\item For $d=3$, the margin is set to 0, and the threshold $\tau$ is chosen such that the total number of type I and II errors is minimized. 
\end{itemize}
The best margin and the corresponding threshold $\tau$ are recorded for each depth. In our implementation, margins are chosen from multiples of 8 in $[8,128]$, and $\tau$ is chosen from multiples of $0.1$ in $[0,1]$.
This offers sufficient granularity with small searching complexity.
If the neighborhood only partially covers a block,
the percentage of the covered area is used as weight to compute the average depth.

The rest 45 frames are fully encoded in $P_2$. 
For a block $B_1$ in $P_1$, $\tilde{X}_2'$ is computed from the encoding results of $P_2$,
using the best margin of the corresponding depth.
Then $\tilde{X}_2'$ is compared to the corresponding $\tau$ of the margin.
if $\tilde{X}_2'<\tau$, $B_1$ remains unpartitioned. In other words, RDO only considers non-split and 2-way splits.
Otherwise, the ordinary RDO is conducted.

\section{Experimental Results}

\begin{table}[t]
\begin{center}
\caption{Encoding results for $\epsilon=0.1$} \label{tab:eps0.1}
\resizebox{0.45\textwidth}{!}{
\begin{tabular}{|l|c|c|c|}
  \hline
  Sequence & BD-rate & BD-PSNR & $\Delta T$
  \\
  \hline
  FourPeople (720p) & 0.51\% & -0.007dB & -33.2\% \\
  Johnny (720p) & 0.90\% & -0.010dB & -35.5\% \\
  Kristen\&Sara (720p) & 0.73\% & -0.010dB & -29.2\% \\
  SlideShow (720p) & 0.70\% & -0.048dB & -23.0\%\\
  BasketballDrive (1080p) & 0.58\% & -0.008dB & -32.5\% \\
  Cactus (1080p) & 0.85\% & -0.013dB & -28.6\% \\
  Kimono (1080p) & 0.73\% & -0.018dB & -26.9\%\\
  ParkScene (1080p) & 0.73\% & -0.018dB & -31.7\% \\
  \hline
  Average & 0.71\% & -0.017dB & -30.1\% \\
  \hline
\end{tabular}}

\caption{Encoding results for $\epsilon=0.2$} \label{tab:eps0.2}
\resizebox{0.45\textwidth}{!}{
\begin{tabular}{|l|c|c|c|}
  \hline
  Sequence & BD-rate & BD-PSNR & $\Delta T$
  \\
  \hline
  FourPeople (720p) & 0.89\% & -0.014dB & -43.7\% \\
  Johnny (720p) & 0.95\% & -0.010dB & -45.1\% \\
  Kristen\&Sara (720p) & 1.14\% & -0.015dB & -39.5\% \\
  SlideShow (720p) & 2.01\% & -0.135dB & -35.0\%\\
  BasketballDrive (1080p) & 0.66\% & -0.010dB & -34.7\% \\
  Cactus (1080p) & 0.95\% & -0.017dB & -31.8\% \\
  Kimono (1080p) & 0.91\% & -0.021dB & -29.7\%\\
  ParkScene (1080p) & 0.84\% & -0.020dB & -34.8\% \\
  \hline
  Average & 1.04\% & -0.030dB & -36.8\%\\
  \hline\end{tabular}}

\caption{Encoding results for algorithm in \cite{Schroeder20152}} \label{tab:old}
\resizebox{0.45\textwidth}{!}{
\begin{tabular}{|l|c|c|c|}
  \hline
  Sequence & BD-rate & BD-PSNR & $\Delta T$
  \\
  \hline
  FourPeople (720p) & 2.80\% & -0.045dB & -42.8\% \\
  Johnny (720p) & 2.00\% & -0.023dB & -46.4\% \\
  Kristen\&Sara (720p) & 2.03\% & -0.029dB & -42.2\% \\
  SlideShow (720p) & 3.37\% & -0.230dB & -31.5\%\\
  BasketballDrive (1080p) & 1.78\% & -0.023dB & -30.7\% \\
  Cactus (1080p) & 1.19\% & -0.020dB & -29.7\% \\
  Kimono (1080p) & 1.95\% & -0.027dB & -31.6\%\\
  ParkScene (1080p) & 1.82\% & -0.048dB & -23.2\% \\
  \hline
  Average & 2.12\% & -0.056dB & -34.7\%\\
  \hline
\end{tabular}}
  
\end{center}
\end{table}

To demonstrate the effectiveness of our inference model,
the fast block structure determination algorithm is integrated into the AV1 encoder to encode 8 test sequences from CTC \cite{CTC},
 with native resolutions of 1280$\times$720 and 1920$\times$1080 (see Table \ref{tab:eps0.1}),
 each consists of 150 frames, using a constant QP of 22, 27, 32, 37,
 the key frame interval is set to 50.
For 720p sequences, they are encoded to target resolutions of 1280$\times$720 and 960$\times$540.
For 1080p sequences, they are encoded to target resolutions of 1920$\times$1080 and 1440$\times$810.

The high target resolution encoding process is accelerated using our proposed algorithm.
The original AV1 encoder is then used to encode the same sequences with the same high target resolutions.
The RD performance and time cost are compared, using BD-rate \cite{BD-rate}, BD-PSNR \cite{BD-psnr}, and the total time cost of high resolution encoding processes of all QP's.
Only time costs of high resolution encoding processes are compared, 
since, as stated before, 
the overall time cost is equal to that of the high resolution encoding process, 
if multiple resolutions encoding are done in parallel.

Two sets of experiments are conducted, 
where the type II error rate threshold $\epsilon$ is set to 0.1 (Table \ref{tab:eps0.1}) and 0.2 (Table \ref{tab:eps0.2}), respectively,
to demonstrate its capability in controlling the trade-off between acceleration and RD performance.
The column $\Delta T$ is the time reduction of our algorithm comparing to the original AV1 encoder.
It is observed that $\epsilon = 0.1$ achieves 30.1\% average time reduction with a negligible 0.71\% BD-rate (0.017dB BD-PSNR loss),
while $\epsilon = 0.2$ achieves a higher average time reduction of 36.8\% but also a higher 1.04\% BD-rate (0.03dB BD-PSNR loss).

For comparison, the latest multiple resolutions encoding algorithm \cite{Schroeder20152} from literature is also evaluated using the same test settings.
Although \cite{Schroeder20152} is based on HEVC and best suited for low resolution encoding acceleration,
it can be migrated to AV1 without difficulty and accelerate high resolution encoding.
Table \ref{tab:old} shows the performance of this algorithm.
Comparing to our proposed algorithm where $\epsilon$ is set to 0.2,
they both achieve about 35\% average time reduction, but the BD-rate (BD-PSNR loss) of our proposed algorithm is halved.
This proves the effectiveness of our proposed algorithm in dealing with high resolution encoding acceleration.

\section{Conclusions}
In this paper, we consider the problem of encoding the same video to different target resolutions using AV1.
We first present the block structure similarity across different resolutions,
and a model based on fineness of frame detail and scale of object motions is proposed to analyze the similarity.
We later see that this model can be used to derive an inference model that 
accelerates high resolution encoding based on low resolution encoding results.
The average block depth of the co-located neighborhood is used to decide early termination in the RDO process.
A bias-variance trade-off can be achieved by searching for an optimal neighborhood range.
Experimental results show that our proposed algorithm offers the capability to control the trade-off between RD performance and time reduction, 
achieving 30.1\%-36.8\% time reduction while keeping BD-rate low at 0.71\%-1.04\%.

\bibliographystyle{IEEEbib}
\bibliography{camera-ready_icme2018template}

\end{document}